\documentclass[aps,prx,amsmath,amssymb,reprint,nofootinbib,longbibliography]{revtex4-1}\usepackage{bbm}
\usepackage{graphicx}
\usepackage{cancel}
\usepackage[caption=false]{subfig}

\usepackage{mdframed}\usepackage{xcolor}

\newcommand\Spp{S_+^+}
\newcommand\Spm{S_+^-}
\newcommand\Smp{S_-^+}
\newcommand\Smm{S_-^-}

\newcommand\Mpp{M_+^+}
\newcommand\Mpm{M_+^-}
\newcommand\Mmp{M_-^+}
\newcommand\Mmm{M_-^-}

\newcommand\barlambda{\bar{\lambda}}
\newcommand\chideg{|\chi|}

\newcommand\rr{\mathbf{r}}

\newcommand\HH{\mathbf{H}}
\newcommand\BB{\mathbf{B}}
\newcommand\DD{\mathbf{D}}
\newcommand\EE{\mathbf{E}}
\newcommand\EEw{\mathbf{E}^\omega(\rr)}
\newcommand\Gpw{\mathbf{G}^\omega_+(\rr)}
\newcommand\Gmw{\mathbf{G}^\omega_-(\rr)}
\newcommand\HHw{\mathbf{H}^\omega(\rr)}
\newcommand\BBw{\mathbf{B}^\omega(\rr)}

\newcommand\ed{\mathbf{d}}
\newcommand\md{\mathbf{m}}

\begin{document}
\title{Objects of maximum electromagnetic chirality}
\author{Ivan Fernandez-Corbaton}\email{ivan.fernandez-corbaton@kit.edu}\affiliation{Institute of Nanotechnology, Karlsruhe Institute of Technology, 76021 Karlsruhe, Germany}
\author{Martin Fruhnert}\affiliation{Institut f\"ur Theoretische Festk\"orperphysik, Karlsruhe Institute of Technology, 76131 Karlsruhe, Germany}
\author{Carsten Rockstuhl}\affiliation{Institut f\"ur Theoretische Festk\"orperphysik, Karlsruhe Institute of Technology, 76131 Karlsruhe, Germany}
\affiliation{Institute of Nanotechnology, Karlsruhe Institute of Technology, 76021 Karlsruhe, Germany}

\begin{abstract}
	We introduce a definition of the electromagnetic chirality of an object and show that it has an upper bound. Reciprocal objects attain the upper bound if and only if they are transparent for all the fields of one polarization handedness (helicity). Additionally, electromagnetic duality symmetry, i.e. helicity preservation upon interaction, turns out to be a necessary condition for reciprocal objects to attain the upper bound. We use these results to provide requirements for the design of such extremal objects. The requirements can be formulated as constraints on the polarizability tensors for dipolar objects or on the material constitutive relations for continuous media. We also outline two applications for objects of maximum electromagnetic chirality: A twofold resonantly enhanced {\em and} background free circular dichroism measurement setup, and angle independent helicity filtering glasses. Finally, we use the theoretically obtained requirements to guide the design of a specific structure, which we then analyze numerically and discuss its performance with respect to maximal electromagnetic chirality.
\end{abstract}
\maketitle

An object is chiral if it cannot be super-imposed onto its mirror image. This simple definition hides significant problems that arise when attempting to measure chirality \cite{Fowler2005}. Quantifying {\em how chiral an object is} is the purpose of scalar measures of chirality which vanish only for achiral objects and assign the same value to an object and its mirror image \cite{Buda1992,Petitjean2003}. There are many different scalar measures of chirality \cite{Petitjean2003}, but none of them allows to sort general objects according to their chirality or to establish what a {\em maximally chiral object} is \cite{Rassat2004} in an unambiguous way.

Independently of these measurement problems, chirality is entrenched in nature: From the lack of mirror symmetry of some interactions among fundamental particles \cite{Wu1957}, to its ubiquitous presence in chemistry and biology. Chirality is studied in very diverse scientific disciplines. One of them is the interaction of chiral matter with
electromagnetic fields, which started two centuries ago \cite{Biot1815} and still attracts significant attention from both its theoretical and practical sides (e.g.
\cite{Lindell1994,Barron2004,Tang2010,Bliokh2011b,Coles2012,Andrews2012,Sersic2012,Coles2013,Canaguier2013,Cameron2014,Bliokh2014,Efrati2014,Nieto2015,Hopkins2015,Poulikakos2016,Papakostas2003,Plum2009,Decker2007,Gansel2009,Decker2010,Zhao2011,Kaschke2012,Khanikaev2012,Schreiber2013,Chaix2013,Schaferling2014,Kruk2014,Shvets2015,Hentschel2015,Plum2015,Jung2015,Asadchy2015,Arteaga2016}). The lack of upper bounds and unambiguous ranking for the magnitude of chirality is a handicap for both theoretical and practical developments. In particular, it is a handicap for the systematic design of chiral structures for interaction with the electromagnetic field. These ambiguities leave us unable to compare different structures and without an extremal reference to design towards. Additionally, it leaves us with no other design guidelines besides chirality itself. We will show that, under a different definition of chirality, chirality upper bounds exist and are attained when objects meet extra requirements. These requirements allow to significantly narrow down the design parameter space.  

In this article, and in the spirit of \cite{NWeinberg2000}, we shift the focus from a geometrical definition of chirality to a definition that is based on the interaction with the field. We introduce a definition of the {\em electromagnetic chirality} of an object based on how it interacts with fields of different polarization handedness (helicity). Our definition can be stated in the following way: An electromagnetically chiral object is one for which all the information obtained from experiments using a fixed incident helicity {\em cannot} be obtained using the opposite one. The various electromagnetic chirality measures arising from this definition take the form of relativistically invariant distances. We then select a particular measure, which can be singled out on physical grounds. We show that the electromagnetic chirality of an object has an upper bound. The upper bound is equal to the square root of the total interaction cross section of the object. Our definition allows the absolute ranking of objects according to their electromagnetic chirality. We show that any object that is transparent to all fields of one helicity attains the upper bound: It is maximally electromagnetically chiral. For reciprocal objects, the implication goes in the other way as well: All maximally electromagnetically chiral and reciprocal objects are transparent to all fields of one helicity. Additionally, we show that any maximally electromagnetically chiral and reciprocal object must have electromagnetic duality symmetry, i.e., interaction shall not change the helicity of the incident fields. We then particularize these results to obtain the constraints that reciprocity plus maximum electromagnetic chirality impose on material constitutive relations, and on the polarizability tensor of an isolated scatterer. These constraints are precise requirements for the design of maximally electromagnetically chiral objects. Electromagnetic duality symmetry is one of them. We then discuss two possible applications for maximally electromagnetically chiral objects: A twofold resonantly enhanced circular dichroism setup, and angle independent helicity filtering glasses. Finally, we numerically analyze the specific design of an object whose properties come close to those of a maximally electromagnetically chiral object in a narrow frequency band. The analysis and results contained in this article apply to linear interactions with finite cross sections. 

\section{Setting}
\begin{figure}[h!]
	\includegraphics[width=0.45\textwidth]{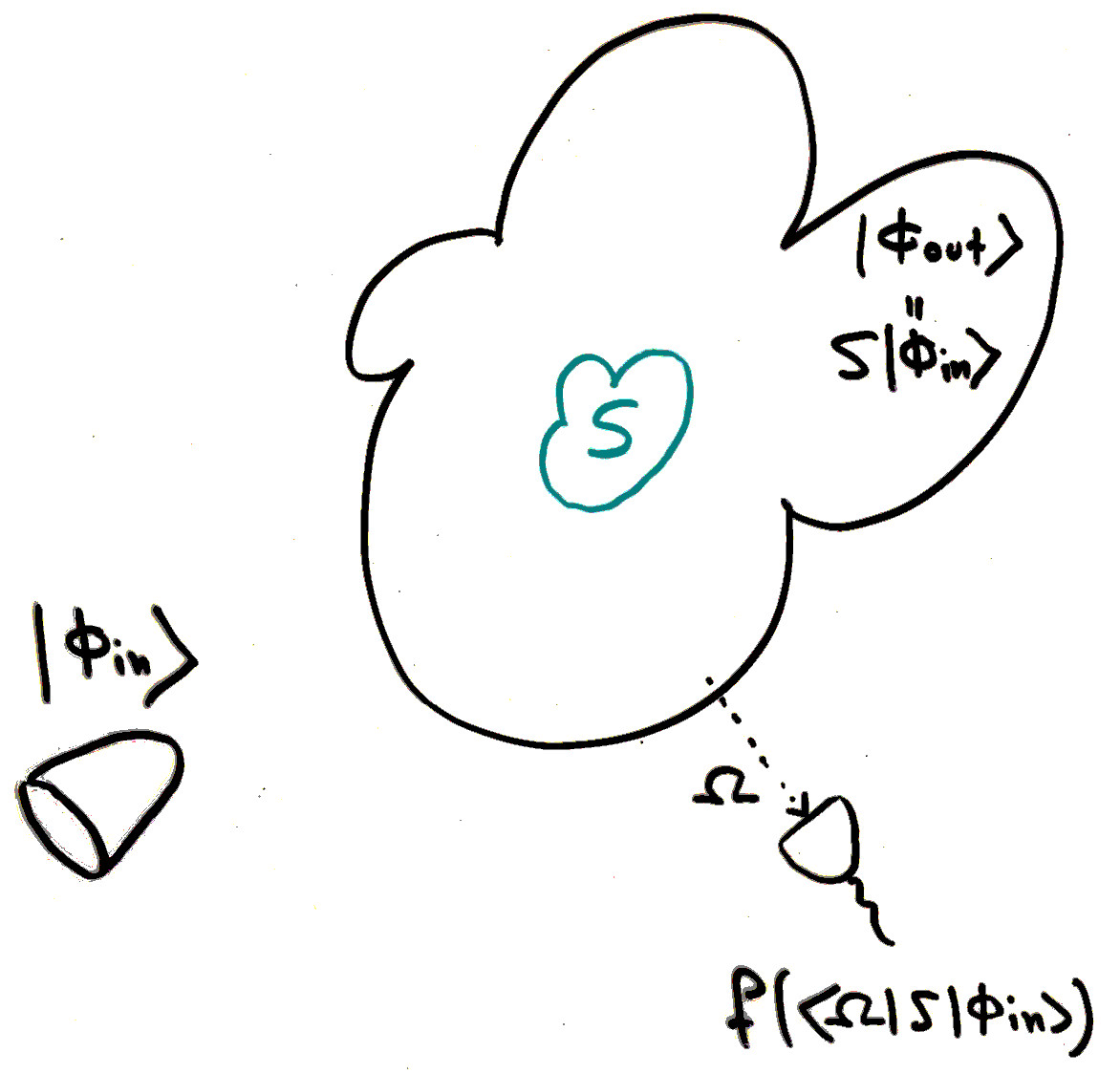}
	\caption{\label{fig:setting} Interaction in the Hilbert space of transverse Maxwell's fields. An incident field $|\Phi_{\text{in}}\rangle$ interacts with an object,  characterized by its interaction operator $S$, and produces a scattered field $|\Phi_{\text{out}}\rangle=S|\Phi_{\text{in}}\rangle$. A detector like the one in the bottom right corner of the figure obtains information about the field scattered through the solid angle $\Omega$: $f\left(\langle\Omega|\Phi_{\text{out}}\rangle\right)=f\left(\langle\Omega|S|\Phi_{\text{in}}\rangle\right)$, where $\langle \Psi|\Gamma\rangle$ is the scalar product of the two vectors $|\Psi\rangle$ and $|\Gamma\rangle$. The interaction operator $S$ also describes absorption and near field illumination and/or measurement.}
\end{figure}

We start with a brief introduction of the setting, and the mathematical tools and notation that we use. The setting is depicted in Fig. \ref{fig:setting}, where an object interacts with an incident field and produces a scattered field and/or absorbs part of the incident energy. We assume linear interaction and finite cross sections. This setting is conveniently treated in the framework of linear operators in Hilbert spaces and Dirac's ``$\langle$bra$|$'' ``$|$ket$\rangle$'' notation \footnote{Dirac introduced this notation in quantum mechanics. It is also very convenient for other situations that can be treated in the framework of Hilbert spaces.}. There, the fields are vectors in the Hilbert space of transverse solutions of Maxwell's equations. The effect of the object is described by its interaction operator $S$, which we take to be the non-trivial part of the scattering operator $\tilde{S}=I+S$  \cite[Sec. 6.4]{Newton2013}. The identity term in $\tilde{S}$ accounts for the portion of the incident field that does not interact with the object, and $S$ is proportional to the system transfer operator $T$ (a.k.a $T$-matrix \cite{Mackowski1996}, \cite[Eq. 2.7.20]{Tsang2000}): $S=2T$. The interaction operator $S$ contains both scattering and absorption information. For example, in Fig. \ref{fig:setting}, a far field detector at solid angular position $\Omega$ provides information about the projection of the scattered field on the corresponding plane wave, i.e. the scattering coefficient $\langle \Omega|S|\Phi_{\text{in}}\rangle$, where $\langle \Omega|$ is a plane wave, and $\langle \Psi|\Gamma\rangle$ the scalar product of $|\Psi\rangle$ and $|\Gamma\rangle$. Besides far field scattering, the interaction operator $S$ also models near field interactions \cite{Zvyagin1998,Quinten1999,Doicu2001}, and absorption by the object. It contains all the information that can be obtained from the object by means of its interaction with transverse electromagnetic fields. We take $S$ as the only relevant representation of the object and define its electromagnetic chirality through the properties of $S$ with respect to the helicity of the fields. The fundamental properties of helicity make it suitable for discussing chiral interactions, as is done in particle physics \cite{Pich2007}.

\section{Using helicity to characterize interactions with chiral matter}
The helicity operator is the projection of the total angular momentum vector operator onto the linear momentum vector operator direction \cite[Sec. 8.4.1]{Tung1985}, \cite[Eq. 57]{Birula2006}, 
\begin{equation}
\Lambda=\frac{\mathbf{J}\cdot{\mathbf{P}}}{|\mathbf{P}|}.
\end{equation}

For classical electromagnetic fields in the complex notation, helicity has two possible eigenvalues $\lambda\in\{1,-1\}$. The eigenstates of helicity are the Riemann-Silberstein linear combinations \cite{Birula1996,Birula2013} $\mathbf{G}_{\pm}=\frac{1}{\sqrt{2}}\left(\mathbf{E}\pm i Z\mathbf{H}\right)$, with $Z$ the medium impedance, so that:
\begin{equation}
	\label{eq:rs}
\Lambda\mathbf{G}_{\pm}=\Lambda \frac{\left(\mathbf{E}\pm i Z\mathbf{H}\right)}{\sqrt{2}}=\pm \frac{\left(\mathbf{E}\pm i Z\mathbf{H}\right)}{\sqrt{2}}=\pm\mathbf{G}_{\pm}.
\end{equation}

Equation (\ref{eq:rs}) can be derived \footnote{See \cite[Eqs.~(2)-(3)]{FerCor2012p}.} from Maxwell's curl equations and the representation of the helicity operator for monochromatic fields of frequency $\omega=kc$, which reads $\Lambda\equiv \frac{\nabla\times}{k}$. Equation (\ref{eq:rs}) is valid in general, including in the near fields around scattering objects. The chiral character of near fields can be readily determined by means of the two Riemann-Silberstein helicity eigenstates. We now show their connection to the optical chirality density \cite{Lipkin1964,Tang2010}. 

We start from a monochromatic electromagnetic field around some scattering object $\left[\EEw,\HHw\right]$. After expressing it in the Riemann-Silberstein basis $\left[\EEw,\HHw\right]\rightarrow\left[\Gpw,\Gmw\right]$, we consider the following space-dependent quantity
\begin{equation}
		\kappa^\omega(\rr)=|\Gpw|^2-|\Gmw|^2,
\end{equation}
that is, the difference between the pointwise intensities of the two helicity eigenvectors. The quantity $\kappa^\omega(\rr)$ completely determines the optical chirality density \cite{Lipkin1964,Tang2010}, defined as
\begin{equation}
\label{eq:C}
C^\omega(\rr)=-\frac{\epsilon\omega}{2} \textrm{Im}\{\EEw^\dagger\BBw\},
\end{equation}
where $\epsilon$ is the permittivity of the surrounding medium. It can be shown \footnote{Equation (\ref{eq:su}) follows from Eq.~(\ref{eq:C}) and these steps:\\ $-\textrm{Im}\{\mathbf{E(\omega)}^\dagger\mathbf{B(\omega)}\}=-\textrm{Im}\{\mathbf{E(\omega)}^\dagger\mu\mathbf{H(\omega)}\}=-\textrm{Im}\{\left(\frac{\mathbf{G_+(\omega)}+\mathbf{G_-(\omega)}}{\sqrt{2}}\right)^\dagger\mu\left(\frac{\mathbf{G_+(\omega)}-\mathbf{G_-(\omega)}}{\sqrt{2}iZ}\right)\}=\frac{1}{2c}\textrm{Im}\{i\left(|\mathbf{G_+(\omega)}|^2-|\mathbf{G_-(\omega)}|^2+2i\textrm{Im}\{\mathbf{G_+(\omega)}^\dagger\mathbf{G_-(\omega)}\}\right)\}=\frac{1}{2c}\left(|\mathbf{G_+(\omega)}|^2-|\mathbf{G_-(\omega)}|^2\right)$} that:
\begin{equation}
		\label{eq:su}
		C^\omega(\rr)=\frac{\epsilon}{4c}\omega \kappa^\omega(\rr).
\end{equation}

The optical chirality density is nowadays widely employed \footnote{A related quantity called helicity density is also used \cite{Cameron2014}. The connections between optical chirality density, helicity density, and the helicity operator have been discussed in the literature \cite{Cameron2012,Bliokh2013}.} to discuss the chiral character of the near fields around scatterers, and the coupling of chiral molecules and dipoles to such near fields (e.g \cite{Tang2010,Canaguier2013,Schaferling2014,Poulikakos2016}). Equation (\ref{eq:su}) confirms the suitability of the helicity formalism for describing chiral near field interactions. 

Helicity can also be understood in operational terms in the momentum (plane wave) representation. An electromagnetic field is an eigenstate of helicity with eigenvalue $+1(-1)$ if and only if all the plane waves in its decomposition are left(right) handed polarized with respect to their corresponding momentum vectors, in which case $\mathbf{G}_-\left(\mathbf{G}_+\right)$ is zero {\em at all points}. The decomposition can contain both propagating and evanescent plane waves \footnote{A plane wave of helicity $+$1($-$1) is the sum(subtraction) of TE and TM plane waves of equal momentum, irrespectively of whether all the momentum components are real or not.}.

For massless fields, the helicity operator commutes with all the transformations of the Poincar\'e group, i.e. space and time translations, spatial rotations, and boosts. It is a relativistic invariant of the field. Additionally, it commutes with the time inversion operator. None of these operations flip the helicity eigenvalues of the states they act on. Crucially, helicity flips only with spatial inversion transformations: $\lambda\rightarrow -\lambda$ after parity, mirror reflections, and rotation-reflections. Helicity is hence a spatial pseudoscalar in the Poincar\'e group extended with space and time inversion.

These properties have already allowed to draw connections between material chirality and optical helicity \cite{Tang2010,Bliokh2011b,Coles2012,Cameron2014}, and to discover the fundamental role of helicity preservation in optical activity \cite{FerCor2012c,FerCor2015,Vidal2015}. 

As an operator, helicity is the generator of the electromagnetic duality transformation \footnote{The duality transformation acts on the initial $(\EE,\HH)$ fields as \cite[Eq. 6.151]{Jackson1998}:\\ $\EE_\theta=\EE\cos\theta  - Z\HH\sin\theta ,$ $Z\HH_\theta=\EE\sin\theta + Z\HH\cos\theta.$}. The relationship between helicity and duality is the same as, for example, angular momentum and rotations. A dual symmetric scatterer preserves the helicity of the fields interacting with it, i.e., it does not couple states of opposite polarization handedness. The conditions for duality symmetry of a scatterer in the macroscopic Maxwell's equations \cite{Lindell2009,FerCor2013} and in the dipolar approximation \cite{Karilainen2012,FerCor2013} are known. The use of helicity and duality for the study and engineering of light matter interactions is developed in detail in \cite{FerCorTHESIS}.

\section{Electromagnetic chirality of an object}
Let us consider the electromagnetic interaction operator $S$ of an object. We choose a basis for transverse Maxwell fields ${|\eta\ \lambda\rangle}$ where helicity is used as the polarization label ($\lambda\in\{1,-1\}$) and $\eta$ is a collective index containing the other three defining numbers \footnote{Each vector of a basis of transverse Maxwell fields, both propagating and evanescent, has four numbers that identify it. These four numbers are the eigenvalues of four commuting operators. For example, multipolar fields are eigenvectors of the angular momentum squared, the angular momentum along one axis, the energy (frequency), and the parity operator. The latter fixes their polarization. The helicity versions of multipolar fields and Bessel beams are the sum and subtraction of the more common parity and TE/TM modes \cite[App. A]{FerCor2012b}. Plane waves can be chosen as eigenstates of the three components of linear momentum, which fixes the frequency, and the helicity operator.}. We can then consider the partial operators $\Spp$, $\Spm$, $\Smp$ and $\Smm$. Each $S_\lambda^{\barlambda}$ acts on input states $|\eta \ \lambda\rangle$ of helicity $\lambda\in\{1,-1\}$, and produces output states $\langle\bar{\lambda}\ \bar{\eta}|$ of helicity $\bar{\lambda}\in\{1,-1\}$.

We define the object to be electromagnetically {\em achiral} if and only if there exist four unitary operators $U_1$, $V_1$, $U_2$ and $V_2$ that commute with the helicity operator, and satisfy
\begin{equation}
	\label{eq:def2}
	\begin{split}
		\langle +\ \bar{\eta}|\Spp|\eta\ +\rangle&=\langle -\ \bar{\eta}|U_1\Smm V_1^{\dagger}|\eta\ -\rangle,\\
		\langle -\ \bar{\eta}|\Spm|\eta\ +\rangle&=\langle +\ \bar{\eta}|U_2\Smp V_2^{\dagger}|\eta\ -\rangle,
	\end{split}
\end{equation}
for all $(\eta,\bar{\eta})$.

Conversely, we define the object to be electromagnetically chiral when its electromagnetic interaction operator never meets Eq.~(\ref{eq:def2}).

Any composition of boosts, rotations, translations, and time inversion is an example of a helicity preserving unitary operator.

We point out that Eq.~(\ref{eq:def2}) says that, for an electromagnetically achiral object, all the information which can be obtained from experiments using only one input helicity can also be obtained from experiments using the opposite helicity. This is not the case for electromagnetically chiral objects.

The common geometrical definition of chirality is a particular case of our definition of electromagnetic chirality. The non-superimposability of an object with its mirror image implies that after $S$ is transformed \footnote{$O\rightarrow X O X^{-1}$ is the transformation rule for an operator $O$ upon the action of operator $X$. The rule for a vector is \unexpanded{$|\Psi\rangle\rightarrow X|\Psi\rangle$}.} by a mirror operator $S\rightarrow MSM^{-1}$, no arbitrary sequence of a rotation $R$ and a translation $T$ can undo the change:
\begin{equation}
	MSM^{-1}\neq (TR) S (TR)^{-1}\text{ for all $T,$ $R$}.
\end{equation}
Conversely, for an achiral object, there exist at least a $TR$ such that
\begin{equation}
	\label{eq:old}
	MSM^{-1}=(TR) S (TR)^{-1}.
\end{equation}
It can be shown \footnote{From Eq.~(\ref{eq:old}) we obtain $S=M^{-1}(TR) S (TR)^{-1}M$, and write $(TR)^{-1}M=(XG)^{-1}$, where $G$ acts only on the polarization index and its action is to flip helicity, and $X$ acts only on the other three indices. Both $G$ and $X$ are unitary. It then follows that: \unexpanded{$\langle +\ \bar{\eta}|\Spp|\eta\ +\rangle=\langle -\ \bar{\eta}|X\Smm X^{\dagger}|\eta\ -\rangle$}, and \unexpanded{$\langle -\ \bar{\eta}|\Spm|\eta\ +\rangle=\langle +\ \bar{\eta}|X\Smp X^{\dagger}|\eta\ -\rangle$}, which is a particular case of Eq.~(\ref{eq:def2}).} that Eq.~(\ref{eq:old}) leads to a particular case of Eq.~(\ref{eq:def2}) with $U_i/V_i$ restricted to rotations and translations. Besides rotations and translations, the proposed definition of electromagnetic chirality allows for other kinds of transformations as well. Notably, the relativistic invariance of electromagnetic helicity allows for $U_i$ and $V_i$ to contain boosts. Consequently, our definition of electromagnetically (a)chiral objects is relativistically invariant. Furthermore, the possibility that $U_i$ and $V_i$ do not represent the same operators is also allowed, and can be interpreted in Eq.~(\ref{eq:def2}) as different input and measurement basis changes.

For the purpose of brevity we will often use the prefix {\em em-} from now on. For example, we will write {\em em-chiral} instead of {\em electromagnetically chiral}.
\subsection{Scalar electromagnetic chirality measures}
The proposed definition has an implication which allows the use of the singular value decomposition to define measures of em-chirality, i.e. measures of {\em how em-chiral an object is}. The singular value decomposition of a complex matrix $A$ always exists, meaning that $A$ can always be written as:
\begin{equation}
	A=BDC^{\dagger},
\end{equation}
where $B$ and $C$ are unitary matrices and $D$ is a diagonal matrix made of real numbers $d_l$ such that $d_l\ge 0$ and $d_1\ge d_2 \ge d_3 \ldots$. The same decomposition exists for completely continuous operators \cite[Chap. II, \S 2]{Gohberg1969}, which can be represented by complex matrices of infinite dimension. The interaction operator $S$ is completely continuous. Our initial assumption of finite cross section guarantees this property \cite[Chap. 8.6]{Weinberg2013}.

Consider the sub-matrices of coefficients
\begin{equation}
	M_\lambda^{\barlambda}\equiv	\langle\barlambda\ \bar{\eta} | S | \eta\ \lambda\rangle\text{ for all }(\eta,\bar{\eta}).
\end{equation}

Let us denote by $\sigma(A)$ the column vector containing the singular values of matrix $A$ in non-increasing order, and define the column vectors 
\begin{equation}
	\label{eq:vpm}
v_+=\begin{bmatrix}\sigma\left(\Mpp\right)\\\sigma\left(\Mpm\right)\end{bmatrix},\
v_-=\begin{bmatrix}\sigma\left(\Mmm\right)\\\sigma\left(\Mmp\right)\end{bmatrix},\
\end{equation}
which contain the singular values of the two sub-matrices corresponding to each input helicity.

The implication of Eq.~(\ref{eq:def2}) for em-achiral objects is that the singular values of $\Mpp$ and $\Mmm$ are equal, and the singular values of $\Mpm$ and $\Mmp$ are equal \footnote{This follows because two matrices ($A,B$) are related by unitary transformations ($U,V$) as $A=UBV^\dagger$, if and only if their singular values are equal \cite[p. 193]{Lancaster1985}.}. This is not the case for em-chiral objects.  The definition of Eq.~(\ref{eq:def2}) is hence equivalent to saying that an object is electromagnetic achiral if and only if $v_+=v_-$. If $v_+ \neq v_-$ the object is electromagnetically chiral. In light of this, any definition of a scalar em-chirality measure $\chideg$ should be based on a distance function between $v_+$ and $v_-$
\begin{equation}
	\label{eq:d}
\chideg= d(v_+,v_-).
\end{equation}
The properties of distance functions ensure that $\chideg$ is real, non negative, and is zero only for em-achiral objects. It is also clear that $\chideg$ is invariant under any transformation by unitary matrices since the singular values remain invariant. The transformations include the matrix representations of translations, rotations, boosts, time inversion, and also parity. The latter flips both the input and output helicities and therefore the two vectors $v_\pm\rightarrow v_\mp$, which, thanks to $d(v_+,v_-)=d(v_-,v_+)$, leaves $\chideg$ unchanged. We conclude that $\chideg$ is relativistically invariant and that it behaves as a scalar chirality measure as defined e.g. in \cite{Buda1992}. We will show in Sec. \ref{sec:mecro} that $\chideg$ is also normalizable to the interval $[0,1]$.

\section{Maximally electromagnetically chiral objects}\label{sec:mecro}
There are many ways of defining the distance between the two vectors in Eq.~(\ref{eq:d}), but there is a physical reason for selecting a particular one.

When an incident state $|\eta\ \lambda\rangle$ interacts with an object and a measurement of the scattering into a different state $\langle \bar{\lambda}\ \bar{\eta}|$ is made, the number of ``clicks'' or the intensities at the detector are proportional to the square of the absolute value of the corresponding coefficient: $|\langle \bar{\lambda}\ \bar{\eta}|S|\eta \ \lambda\rangle|^2$. Let us consider the sum over all possible incident and output states 
\begin{equation}
	C_{int}=\sum_{\lambda\bar{\lambda}}\sum_{\eta\bar{\eta}} |\langle \bar{\lambda}\ \bar{\eta}|S|\eta \ \lambda\rangle|^2.
\end{equation}
This quantity can be understood as the total interaction cross section of the object. It is a measure of the overall coupling between the object and the electromagnetic field, including both scattering and absorption. It can be shown \footnote{The total sum can be computed by adding the four partial sums for each combination of incident and output helicities \unexpanded{$C_{int}=\sum_{\lambda\bar{\lambda}}C_{int}^{\lambda\bar{\lambda}}=\sum_{\lambda\bar{\lambda}}\sum_{\eta\bar{\eta}} |\langle \bar{\lambda}\ \bar{\eta}|S|\eta \ \lambda\rangle|^2=\sum_{\lambda\bar{\lambda}}\mathrm{trace}\left({M_{\lambda}^{\bar{\lambda}}}^\dagger M_{\lambda}^{\bar{\lambda}}\right)}$. The final result $C_{int}=(v_+)^Tv_++(v_-)^Tv_-$ is reached using Eq.~(\ref{eq:vpm}) and the properties of the singular value decomposition.} that
\begin{equation}
	\label{eq:cint}
	C_{int}=(v_+)^Tv_++(v_-)^Tv_-, 
\end{equation}
where $^T$ means transposition. The total interaction cross section is the sum of the interaction cross sections that the object presents to each input helicity:
\begin{equation}
	\label{eq:cintlambda}
	C_{int}^+=(v_+)^Tv_+,\ C_{int}^-=(v_-)^Tv_-. 
\end{equation}

We see that the total interaction cross section is the sum of the squared Euclidean norms of $v_+$ and $v_-$.  We hence select the Euclidean norm \footnote{A norm is always a distance. The converse is not true.} to compute $\chideg$: 
\begin{equation}
	\label{eq:chiipart}
	\chideg=\sqrt{\left(v_+-v_-\right)^T\left(v_+-v_-\right)}=\sqrt{\sum_l (v_+(l)-v_-(l))^2}.
\end{equation}

Since a particular norm can be chosen on physical grounds, we can establish an absolute ordering of objects with respect to their em-chirality using Eq.~(\ref{eq:chiipart}).

We now reach some notable results: With the definition of Eq.~(\ref{eq:chiipart}), the electromagnetic chirality of an object is upper bounded. All objects that are transparent for fields of one helicity achieve the upper bound: They are maximally electromagnetically chiral. If the object is reciprocal, the implication goes also the other way: All maximally em-chiral reciprocal objects are transparent to fields of one helicity. Additionally, all maximally em-chiral reciprocal objects preserve helicity upon interaction, i.e. have electromagnetic duality symmetry.

In order to show all this, we start by fixing a given $C_{int}\neq 0$, and compute the ratio $\chideg^2/C_{int}$:
\begin{equation}
	\label{eq:mostchiral}
	\frac{\chideg^2}{C_{int}}=\frac{ (v_+)^Tv_++(v_-)^Tv_-- 2(v_+)^Tv_-}{C_{int}}= 1- 2\frac{(v_+)^Tv_-}{C_{int}}.\\
\end{equation}

Since the elements of $v_\pm$ are all real and non negative, the term $(v_+)^Tv_-$ is always greater or equal than zero. It follows that $\chideg$ is upper bounded by $\sqrt{C_{int}}$:
\begin{equation}
\chideg\in\left[0,\sqrt{C_{int}}\right].
\end{equation}
It also follows that the upper bound is attained if and only if $(v_+)^Tv_-=0$, which means [Eq. (\ref{eq:vpm})]:
\begin{equation}
	\label{eq:zero}
	0=\sigma\left(\Mpp\right)^T\sigma\left(\Mmm\right)+\sigma\left(\Mpm\right)^T\sigma\left(\Mmp\right).
\end{equation}
We now exploit that the elements of $\sigma(A)$ are real, greater or equal than zero, and {\em sorted in non-increasing order} to conclude that: a) Equation (\ref{eq:zero}) is only met when both terms in the sum are simultaneously zero, because each individual term is greater or equal than zero, and, b) let us assume that a term is zero: This means that at least one of the two matrices involved must be null because at least one of the two involved vectors of singular values, whose elements are non-increasing, must contain only zeros.

Objects that are transparent to one of the helicities of the field always meet Eq. (\ref{eq:zero}), and are hence always maximally em-chiral. This is clear from the conditions of transparency to one helicity: Either $\Mpp=\Mpm=0$ or $\Mmm=\Mmp=0$.

The converse is not necessarily true. The following two cases meet Eq. (\ref{eq:zero}) but are not transparent to one helicity:
\begin{equation}
	\label{eq:nonrec}
	\begin{split}
		&\left(\Mpp=0,\Mpm\neq0,\Mmm\neq0,\Mmp=0\right),\\
		&\left(\Mpp\neq0,\Mpm=0,\Mmm=0,\Mmp\neq0\right).\\
	\end{split}
\end{equation}
Using a) and b) above it is easy to see that these are the only cases of maximally em-chiral objects that are not transparent to one helicity. We now show that both cases violate reciprocity, and hence, that any reciprocal maximally em-chiral object must be necessarily transparent to one helicity.

In the basis of plane waves with well defined momentum $\mathbf{p}$ and helicity $\lambda$, the reciprocity condition \cite[Eq. 2.22]{Sapienza2005} results in the following relationships between input and output states \footnote{The reciprocity condition is given in \cite[Eq. 2.22]{Sapienza2005}: \unexpanded{$\langle \varepsilon_f\ \mathbf{p}_f|S|\mathbf{p}_i\ \varepsilon_i\rangle=\langle \varepsilon_i^*\ {-\mathbf{p}_i}|S|{-\mathbf{p}_f}\ \varepsilon_f^*\rangle$}, where $\varepsilon_{f,i}$ are general polarization vectors. Equation (\ref{eq:rec}) follows from using helicity polarization vectors $\varepsilon(\mathbf{p},\lambda)$, which are the sum and subtraction of the TE and TM polarization vectors, and the correspondences with our notation $\unexpanded{|\mathbf{p} \ \varepsilon(\mathbf{p},\lambda) \rangle\equiv |\mathbf{p}\ \lambda\rangle$} and \cite[Sec. 3]{Birula2013}: \unexpanded{$|{-\mathbf{p}} \ \varepsilon^*(\mathbf{p},\lambda) \rangle=|{-\mathbf{p}}\ \varepsilon({-\mathbf{p}},\lambda) \rangle\equiv |{-\mathbf{p}}\ \lambda\rangle$}.}:
	\begin{equation}
	\label{eq:rec}
	\langle \bar{\lambda}\ {\mathbf{\bar{p}}}|S|\mathbf{p} \ \lambda\rangle=
	\langle {\lambda}\ {-\mathbf{p}}|S|{-\mathbf{\bar{p}}} \ \bar{\lambda} \rangle.
\end{equation}
Equation (\ref{eq:rec}) means in particular that 
\begin{equation}
	\langle +\ {\mathbf{\bar{p}}}|S|\mathbf{p} \ -\rangle=
	\langle -\ {-\mathbf{p}}|S|{-\mathbf{\bar{p}}} \ +\rangle.
\end{equation}
Therefore, if the object is reciprocal, $\Mpm=0\iff\Mmp=0$, which holds independently of the choice of basis. The two cases in Eq. (\ref{eq:nonrec}) violate this condition and must hence be non-reciprocal.

Finally, we observe that reciprocal maximally em-chiral objects must meet $\Mpm=\Mmp=0$. This condition is the definition of helicity preservation and is equivalent to the statement that the object has electromagnetic duality symmetry. We have reached the conclusion that all maximally em-chiral reciprocal objects are necessarily dual symmetric. Duality is hence a requirement for reciprocal objects to be maximally em-chiral objects. 

We have proved all the previously announced results, which we summarize here. 
\begin{mdframed}[backgroundcolor=gray!20] 
	\vspace{0.25cm}
For reciprocal and non-reciprocal objects:
\begin{equation}
	\nonumber
	\begin{split}
	&\chideg\in\left[0,\sqrt{C_{int}}\right],\\
	&\text{Transparency to one helicity}\implies \chideg=\sqrt{C_{int}}.
	\end{split}
\end{equation}
For reciprocal objects:
\begin{equation}
	\nonumber
	\begin{split}
		&\text{Transparency to one helicity}\iff\chideg=\sqrt{C_{int}},\\
		&\chideg=\sqrt{C_{int}}\implies\text{Duality symmetry}.
	\end{split}
\end{equation}
	\vspace{0.05cm}
\end{mdframed}

It is worth mentioning that reciprocal interaction does not need to be lossless, and that when it is, time reversal invariance is automatically fulfilled. These results have a notable parallelism with portions of the chiral electroweak theory in the standard model of high energy physics, where only left chiral fermions interact via the weak force, the interaction is unitary (lossless) and time reversal is a good symmetry \cite[Sec. 3.3.1]{Pich2007}. 

When there are material losses, the scattering operator $\tilde{S}=I+S$ is not unitary. The total absorption cross section for each input helicity $\lambda$ can be computed as: 
\begin{equation}
	\label{eq:losses}
\text{trace}\left(I-L_\lambda^\dagger L_\lambda\right), \text{ where } L_{\lambda}=\begin{pmatrix}I+M_\lambda^\lambda\\M_\lambda^{-\lambda}\end{pmatrix}.
\end{equation}

\begin{figure}[h!]
	\subfloat[General object. Interacts with and mixes both helicities.]{\includegraphics[width=\linewidth]{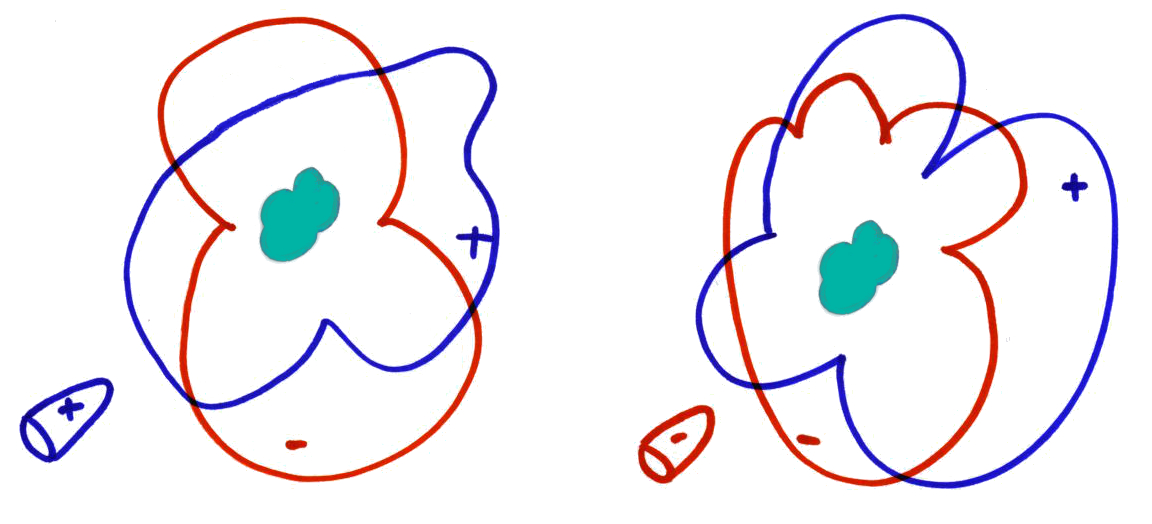}}\\
		\subfloat[Maximal electromagnetically chiral and reciprocal object. It is transparent to one helicity and preserves helicity upon interaction.]{\includegraphics[width=\linewidth]{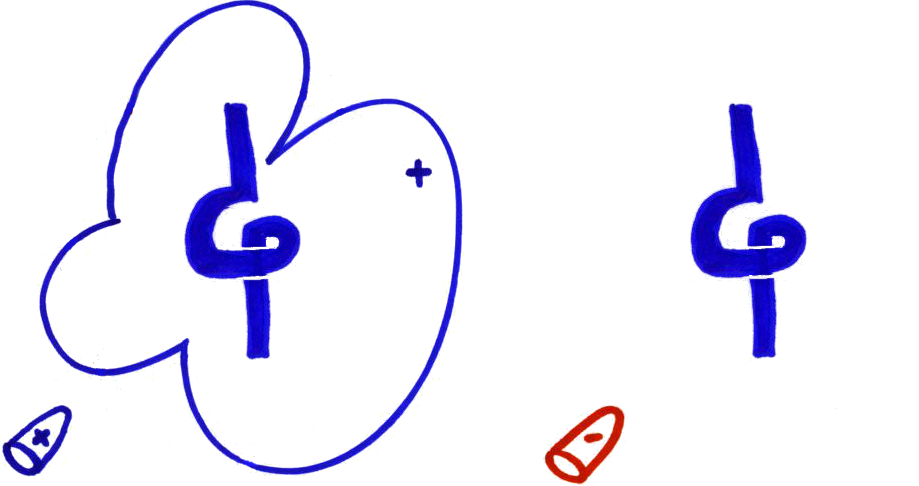}}
	\caption{\label{fig:classes} Interaction of a general object (a), and a reciprocal maximally em-chiral object (b) with fields of pure helicity $\pm$1. Fields of helicity +1 are blue and marked with a ``+''. Fields of helicity $-$1 are red an marked with a ``$-$''. Incoming fields are drawn as bullet-like shapes and scattered fields as clouds surrounding the scatterers. A general object interacts with and mixes both helicities (a). For reciprocal objects (b), maximal electromagnetic chirality occurs if and only if the object is transparent to one helicity. It also implies that the object preserves helicity upon interaction: The scattered field shall have the same helicity as the incident field. The object must hence have electromagnetic duality symmetry.}
\end{figure}

Figure \ref{fig:classes} depicts the different behavior that a general object and a reciprocal maximally em-chiral object have with respect to their interaction with fields of pure helicity.

We will now discuss reciprocity together with maximum em-chirality in the macroscopic equations and in the dipolar approximation. 

\subsection{Constraints in polarizability tensors and constitutive relations}\label{sec:constraints}
For an object embedded in an isotropic and homogeneous medium with permittivity and permeability $(\epsilon_s,\mu_s)$, the conditions of transparency to one helicity and reciprocity restrict the constitutive relations of the material of which the object is made
\begin{equation}
\label{eq:cons}
\begin{bmatrix}Z_s\DD\\\BB\end{bmatrix}=\begin{bmatrix}\epsilon&\chi\\\gamma&\mu\end{bmatrix}\begin{bmatrix}\EE\\Z_s\HH\end{bmatrix},
\end{equation}
 through the relations 
 \begin{equation}
	 \label{eq:three}
	 \begin{split}
		 \text{reciprocity:$^{\text{\cite[Eq. 5.5-17 ]{Kong1990}}}$ }& \epsilon=\epsilon^T,\ \text{  } \mu=\mu^T,\ \text{  } \chi^T=-\gamma,\\
		 \text{transparency to helicity +1: }&\boxed{\epsilon=i\chi,\ \text{   }\mu=-i\gamma,}\\
		 \text{transparency to helicity $-$1: }&\boxed{\epsilon=-i\chi,\ \text{   }\mu=i\gamma,}\\
	 \end{split}
\end{equation}
Each of the $(\epsilon,\mu,\chi,\gamma)$ is a 3$\times$3 tensor. The boxed equations for transparency to one helicity are readily reached by changing the basis in Eq.~(\ref{eq:cons}) to the combinations $\mathbf{F}_{\pm}=\frac{1}{\sqrt{2}}\left(Z_s\DD\pm i\BB\right)$ and $\mathbf{G}_{\pm}=\frac{1}{\sqrt{2}}\left(\mathbf{E}\pm iZ_s\HH\right)$ and nulling the appropriate column of $3\times3$ blocks for transparency to the +1 or -1 helicity, namely 
\begin{equation}
		\label{eq:last}
\begin{bmatrix}\mathbf{F}_+\\\mathbf{F}_-\end{bmatrix}=\begin{bmatrix}0&b\\0&a\end{bmatrix}\begin{bmatrix}\mathbf{G}_+\\\mathbf{G}_-\end{bmatrix},\text{ or }\
\begin{bmatrix}\mathbf{F}_+\\\mathbf{F}_-\end{bmatrix}=\begin{bmatrix}\bar{a}&0\\\bar{b}&0\end{bmatrix}\begin{bmatrix}\mathbf{G}_+\\\mathbf{G}_-\end{bmatrix}.
\end{equation}

As expected, the first line in Eq.~(\ref{eq:three}) plus any of the other two imply duality symmetry \cite{Lindell2009,FerCor2013}: $\epsilon=\mu,\ \chi=-\gamma$. This forces $b$ and $\bar{b}$ in Eq. (\ref{eq:last}) to be equal to zero. In the end, the only freedom left in a maximally em-chiral reciprocal object is a symmetric three by three complex tensor and the choice of transparency to the +1 (upper signs) or $-$1 (lower signs) helicity: 
\begin{equation}
 \label{eq:mcrcons}
	\epsilon=\epsilon^T=\mu=\pm i\chi=\mp i\gamma.
\end{equation}

In the field of metamaterials, effective constitutive relations are obtained from the joint response of an ensemble of electromagnetically small objects. The response of a small enough object is approximately determined by its induced electric ($\ed$) and magnetic ($\md$) dipolar response
\begin{equation}
\begin{bmatrix}\ed\\\md\end{bmatrix}=\begin{bmatrix}\alpha_{dE}& \alpha_{dH}\\ \alpha_{mE} & \alpha_{mH}\end{bmatrix}\begin{bmatrix}\EE\\\HH\end{bmatrix}.
\end{equation}

The same kind of analysis that lead us to Eq.~(\ref{eq:mcrcons}) leads to a similar result. The reciprocity conditions for polarizability tensors have the same form as in Eq.~(\ref{eq:three}) \cite{Sersic2011}. Transparency to one helicity can be imposed by changing the fields as before and changing the dipoles to the combinations $\left(\ed\pm i\md/c\right)/\sqrt{2}$. These combinations radiate fields of single helicity content \cite[Sec. 2.4.3]{FerCorTHESIS}. Again, reciprocity plus transparency to one helicity imply (dipolar) duality ($\alpha_{dE}=\epsilon_s\alpha_{mH},\ \alpha_{mE}=-\alpha_{dH}/\mu_s$), and the final result is
\begin{equation}
	\label{eq:mcrdipolar}
	\alpha_{dE}=\alpha_{dE}^T=\epsilon_s\alpha_{mH}=\pm i \alpha_{dH}/Z_s=\mp i\mu_s\alpha_{mE}/Z_s.
\end{equation}
We note that the findings in \cite{Sersic2012}, obtained for the particular case of planar circuits, are consistent with our results.

The conditions in Eq.~(\ref{eq:mcrdipolar}) describe maximally em-chiral dipolar objects which do not couple to one of the helicity components of the field $\mathbf{G}_{\pm}$. This zero coupling is independent of whether $\left(\mathbf{E},\mathbf{H}\right)$ are far fields in the radiation zone or near fields around a scatterer.
 
\section{Applications}\label{sec:applications}
Before discussing two practical applications of maximally em-chiral reciprocal objects, we highlight two remarkable benefits of using helicity to treat the polarization of the field \cite[Sec. 2.9]{FerCorTHESIS}, which we will exploit.

\begin{figure*}[ht!]
	\subfloat[Illumination with a beam of helicity +1, which excites the blue object on the left.]{\includegraphics[width=0.25\linewidth]{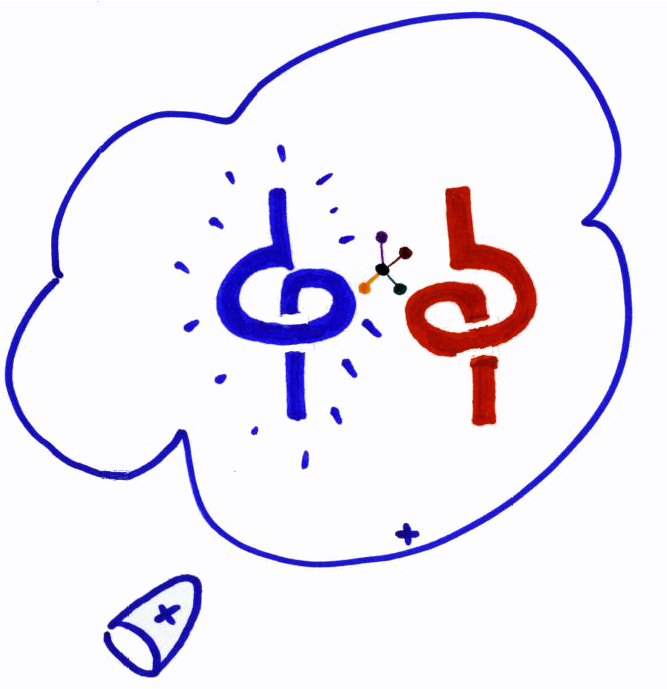}}\hspace{0.35cm}
	\subfloat[(Zoomed) The molecule is illuminated by the near field of the excited object, which is of helicity +1, and produces a weak field with both helicities.]{\raisebox{0.2\height}{\includegraphics[angle=90,width=0.35\linewidth]{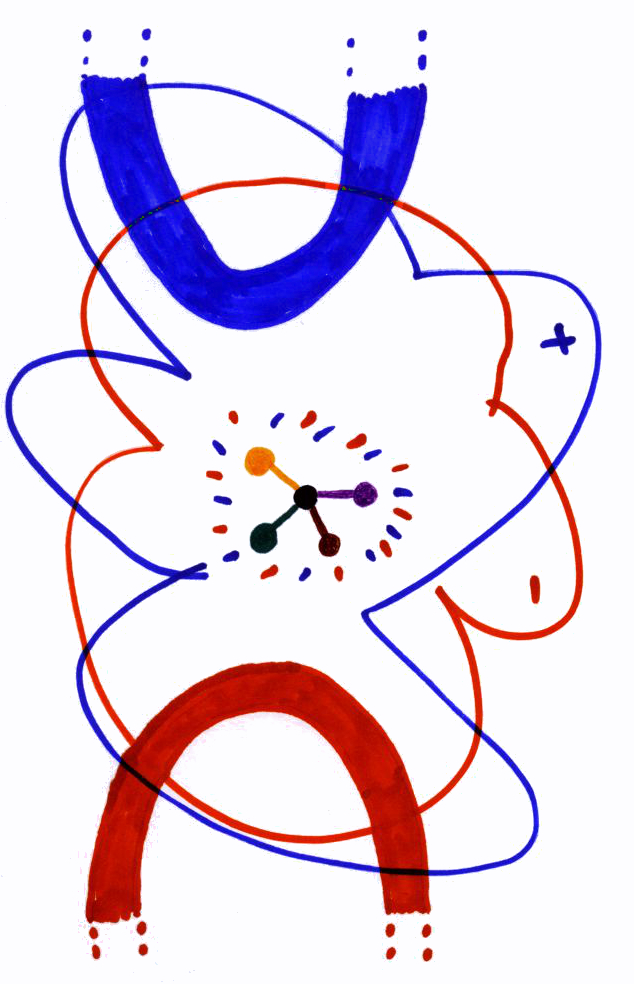}}}\hspace{0.35cm}
	\subfloat[The molecular field with helicity -1 excites the red object on the right, which produces a strong field of helicity -1. This field is measured.]{\raisebox{0\height}{\includegraphics[width=0.35\linewidth]{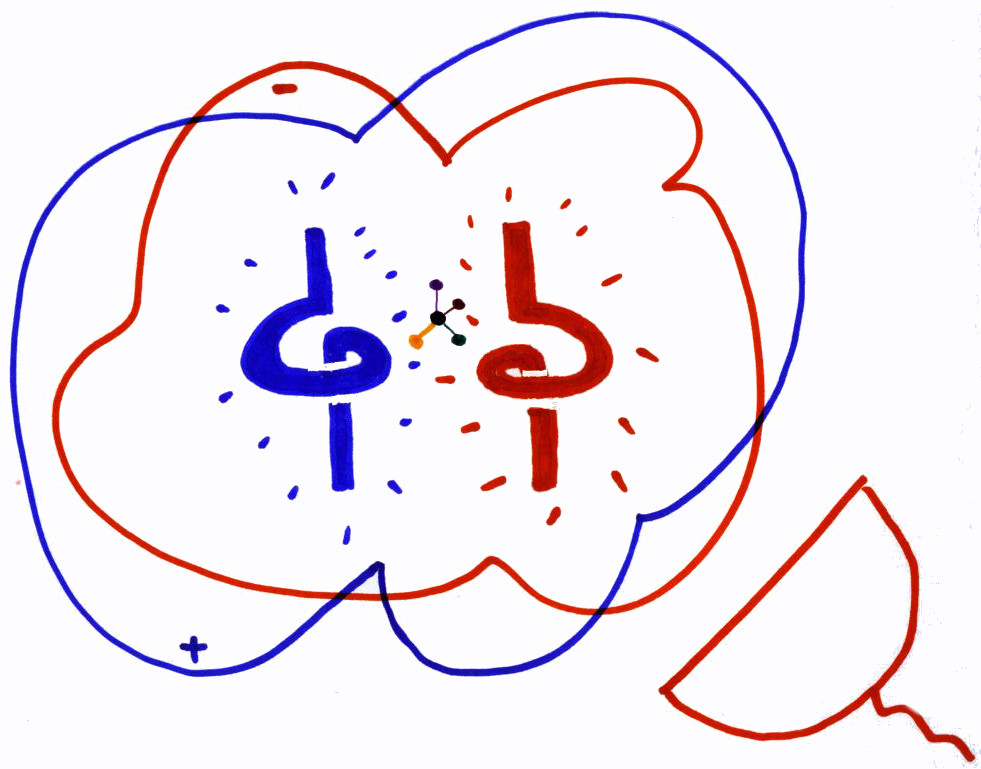}}}
	\caption{\label{fig:cd} Double resonant enhancement for circular dichroism measurements. Panel (a): Two resonant maximally electromagnetically chiral and reciprocal objects are placed close to each other. A chiral molecule is in their vicinity. The external illumination excites only one of the objects, whose resonance illuminates the molecule with a strong field of the same helicity as the incident beam. Panel (b): Upon illumination, the molecule produces a weak scattered field containing both helicities, which excite the two resonant objects. Panel (c): The scattered field of helicity opposite to the initial one is measured. This field exists due to the presence of the molecule. The other half of the circular dichroism measurement is obtained by interchanging the input and measured helicities. The final difference features the two resonant enhancements of opposite helicity: One in illumination and one in the amplification of the field scattered by the molecule.} 

\end{figure*}

First, helicity commutes with rotations and translations. This means that after rotating and displacing a helicity preserving object, it remains helicity preserving. This is not the case if one uses a different description of the polarization. For example, an object with parity inversion symmetry, like a sphere, preserves the parity of the fields interacting with it when located in the origin of coordinates. After a displacement, the multipoles of different parity will mix with each other upon scattering. 

Second, helicity preservation and transparency to one helicity are properties which do not depend on whether one considers the near, intermediate, or far field zones. At the root of this property lies the fact that, for a field of pure helicity one of the two combinations $\mathbf{E}\pm iZ \mathbf{H}$ in Eq.~(\ref{eq:rs}) is equal to zero {\em at all spacetime points}. For example, the field scattered off a dual symmetric object upon illumination with a general field of helicity $\lambda$ ($\mathbf{G}_\lambda\neq 0$) has zero component of helicity $-\lambda$ ($\mathbf{G}_{-\lambda}=0$) in any field zone. This is illustrated in \cite[Fig. 1]{FerCor2012p} for the near fields. Different dual symmetric objects are illuminated with a field of $\lambda=+1$. The numerical solution of Maxwell's equations show that the fields at a distance of about 1/30-$th$ of the wavelength away from the objects have zero intensity of the $\lambda=-1$ component. Similarly, an otherwise arbitrarily complex near field of pure helicity will not excite an object which is transparent to that helicity. This can be deduced from the constraints for transparency and duality symmetry in Sec. \ref{sec:constraints}. When expressed in the helicity basis, the constitutive relations and polarizability tensors for maximal em-chirality have a 3$\times$3 block structure like:
\begin{equation}
\begin{bmatrix}0&0\\0&a\end{bmatrix}\text{, or }\begin{bmatrix}\bar{a}&0\\0&0\end{bmatrix},
\end{equation}
which ensures a null response to $\mathbf{E}-iZ\mathbf{H}$ or $\mathbf{E}+iZ\mathbf{H}$, respectively. The null response is independent of whether the $\mathbf{E}$ and $\mathbf{H}$ fields belong to the near, intermediate, or far field zones of the exciting source.

We now sketch two concept proposals for applications of maximally em-chiral and reciprocal objects: Enhanced circular dichroism measurements of molecules and angle independent helicity filtering glasses.

\subsection{Double resonantly enhanced circular dichroism setup}

Circular dichroism (CD) is used to distinguish between the two enantiomeric forms of chiral molecules. This distinction is particularly important for synthetic drug production because the two enantiomers can have very different effects. The weak response of the molecules typically results in low sensitivity and/or long measurement times. Geometrically chiral plasmonic structures featuring strong scattered near fields upon external illumination are being studied for enhancing the CD signal of the molecules in their vicinity. This design principle has two important drawbacks. One is that the near field of a general geometrically chiral structure is not of pure helicity, even when the external excitation is (see Fig. \ref{fig:classes}(a)). The molecule is thus illuminated by a field of mixed handedness which blurs the CD measurement. The second drawback is that the plasmonic structure itself produces a strong CD signal. We argue that a double resonantly enhanced circular dichroism setup can be designed by placing two resonant maximally em-chiral reciprocal objects of opposite handedness close to each other, and that this scheme avoids the two aforementioned problems. 

Let us start by considering two maximally em-chiral reciprocal objects of opposite handedness $O$ and $\bar{O}$. Straightforward symmetry arguments show that if $O$ is a maximally em-chiral reciprocal object with a resonance for helicity +1, a suitable $\bar{O}$ can be obtained as the mirror image of $O$, which will be a maximally em-chiral reciprocal object with a resonance at the same frequency as $O$, but for the opposite helicity. 

As previously discussed, if we place $O$ and $\bar{O}$ close together, they remain electromagnetically uncoupled, independently of their relative orientation or separation. As a result, illuminating the pair with light of a given helicity does not produce any scattering of the opposite helicity. In Fig. \ref{fig:cd}, a chiral molecule is in the vicinity of such a system. The three panels show a sequence of events for illustration purposes. In Fig. \ref{fig:cd}(a) an external field of well defined helicity $\lambda=1$ is incident on the system. The resonance in $O$ will illuminate the molecule with a strong field of pure helicity $\lambda=1$. Assuming that the molecule is not dual symmetric \footnote{Duality symmetry requires a molecule to have comparable electric and magnetic responses, which is not the case for most chiral molecules.}, the interaction will result in a weak field containing both helicities, as depicted in Fig. \ref{fig:cd}(b). The molecular field will excite both structures. In particular, the portion with $\lambda=-1$ will excite the resonance of $\bar{O}$ producing a strong scattered field of $\lambda=-1$ that can then be measured by an apparatus which selects a single field handedness (Fig. \ref{fig:cd}(c)). The measured power of the $\lambda=-1$ component depends on the helicity flipping operator $\Spm$ of the molecule. We note that the measurement has been enhanced by two resonant interactions of opposite helicity, one in amplifying the illumination and one in amplifying the field scattered by the molecule. The other half of the circular dichroism measurement is obtained by changing the helicity of the incident field and measurement apparatus. Chiral molecules have $\Spm\neq \Smp$ where the difference depends on the magneto-electric part of their polarizability tensors. The difference between the two measurements will feature the twofold enhancement. The scheme is suitable for distinguishing between the two enantiomeric forms of a chiral molecule.

Finally, we note that the generation and measurement of pure helicity modes in the collimated regime at optical frequencies is straightforward and can be done with polarizers and quarter wave-plates \cite{FerCor2012b,Tischler2014}, and that microscope objectives designed to meet the aplanatic approximation preserve helicity \cite[App. C]{FerCor2012b}, which makes them suitable as focusing and collecting lenses in the proposed measurement scheme. 
\subsection{Angle independent helicity filtering glasses}
\begin{figure}[h!]
	\includegraphics[width=\linewidth]{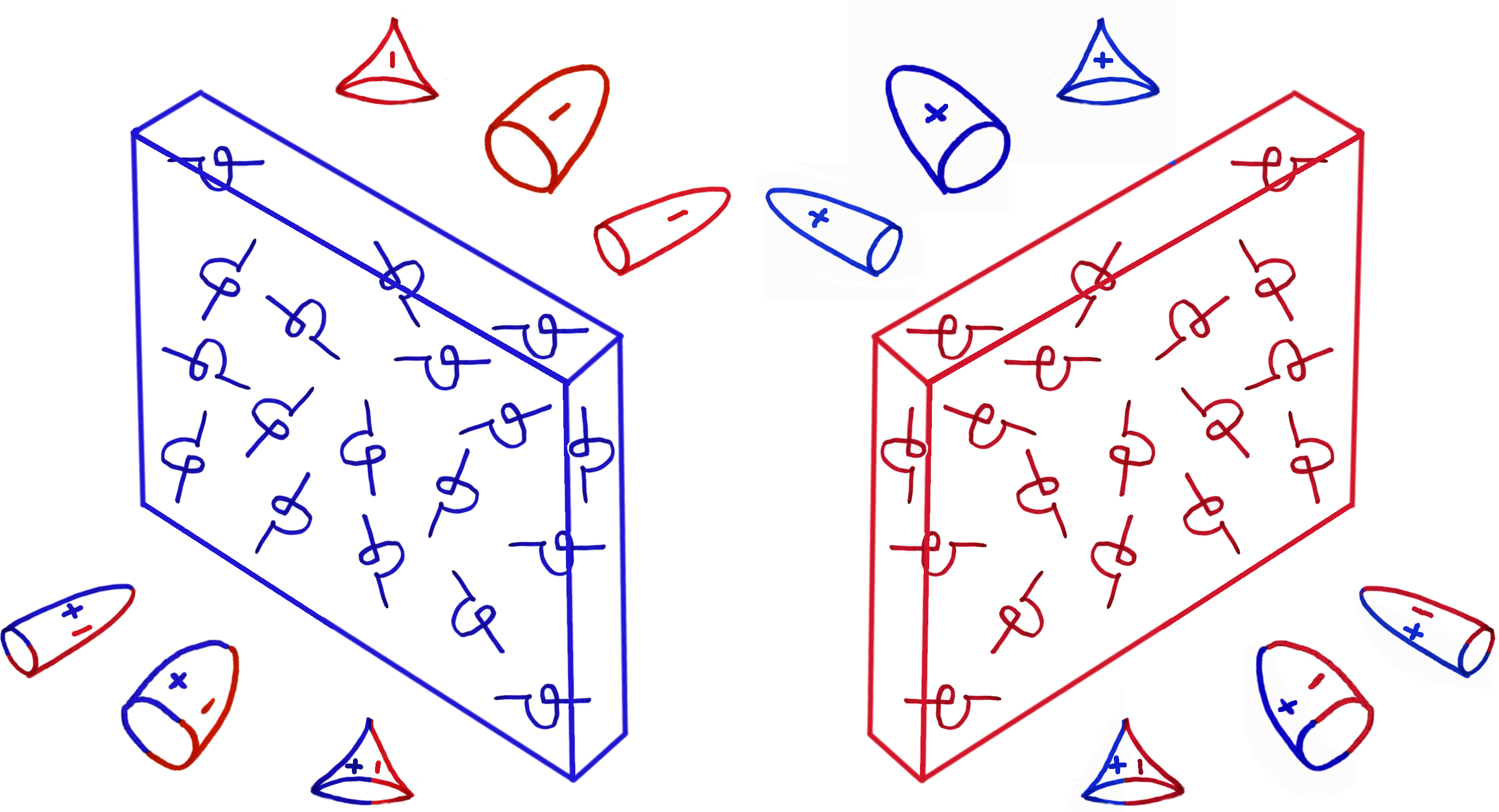}
	\caption{\label{fig:glasses} Two slabs containing lossy maximally electromagnetically chiral and reciprocal objects of opposite handedness. For large enough slab thickness/particle density/losses, each slab filters out one of the helicities by absorption. The other one passes right through. This behavior is independent of the angle of incidence. The slabs can be used to design glasses for viewing 3D projections where the images destined for each eye are encoded in the two circular polarizations. The glasses would allow to see the 3D effect even at large angles from the perpendicular of the projector.}
\end{figure}

A second application is helicity filtering glasses. For this purpose we consider a slab of material containing randomly arranged maximally em-chiral and reciprocal particles with losses. For large enough slab thickness/particle density/losses, the slab will filter out one of the helicities by absorption. The other helicity will pass straight through. This behavior is independent of the angle of incident of the field due to the orientation independent character of helicity preservation and transparency. Two of these slabs made with particles that are the mirror image of each other make suitable glasses for viewing 3D projections where the images destined for each eye are encoded in the two circular polarizations (see Fig. \ref{fig:glasses}). The filtering ability of the glasses is independent of the relative orientation between the user and the projector. This is in sharp contrast to designs based on the paraxial optical paradigm of ``quarter wave plate plus linear polarizer'', whose polarization discrimination degrades as the angle of incidence deviates from the normal. 

\section{Numerical study}
\begin{figure}[ht]
\includegraphics{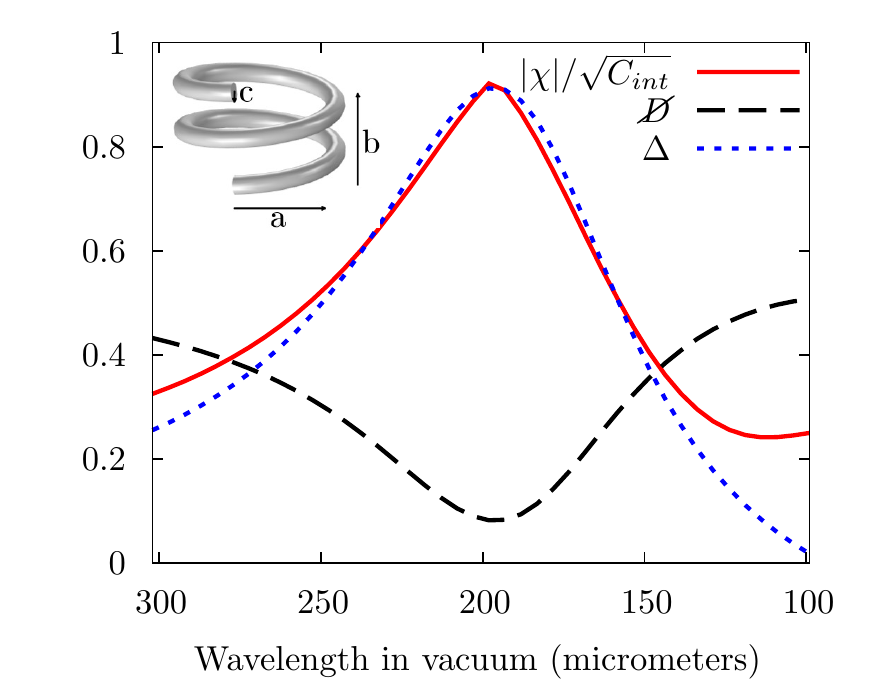}
\caption{\label{fig:chifixed}Normalized em-chirality $\chi/\sqrt{C_{int}}$, measure of duality breaking $\cancel{D}$, and contrast of helicity cross sections $\Delta$ for the double turn silver helix shown in the inset. Large values of $\chi/\sqrt{C_{int}}$ coincide with simultaneously large values of $\Delta$ and small values of $\cancel{D}$. The dimensions of the helix are: Major radius $a=6.48$ $\mu$m, height $b=8.52$ $\mu$m, and wire radius $c=0.8$ $\mu$m.}
\end{figure}

In this final section, we study the em-chirality properties of the double turn silver helix depicted in the inset of Fig. \ref{fig:chifixed}. The aim of the study is two fold. On the one hand we use a realistic object to illustrate two important ideas: Large em-chirality needs duality symmetry, and, large em-chirality is possible in the presence of material absorption. On the other hand, we show a realistic object which, in a narrow frequency band, approaches the maximum em-chirality. 

Our choice of structure is motivated by the geometrically optimized helical antennas for circular polarization \cite{Wheeler1947}. Notable properties of similar antennas have been recently studied \cite{Semchenko2009,Karilainen2012b,Radi2013}. Under some approximations, the geometrically optimized helical antennas can be shown to meet the dipolar duality condition at their resonance frequency. Additionally, they present largely different cross sections to the two polarization handedness of plane waves with momentum perpendicular to the helix axis. Instead of using common approximations like thin helix wire or restriction to dipolar scattering, we obtain the complete interaction matrix at each frequency by using the permittivity of silver from Ref. \onlinecite{Hagemann1975} and a technique similar to the one described in \cite[Sec. 4.1]{Gimbutas2013}. Exact numerical solutions of the Maxwell's equations based on a finite-element method allow us to obtain the T-matrix of the helix, which is related to the interaction operator \cite[Eq. 2.7.20]{Tsang2000} as $S=2T$. To obtain the geometrical parameters of the helix, a first initial guess using Ref. \onlinecite{Wheeler1947} is made. This is followed by a local tuning of its major radius $a$ and height $b$. The local tuning seeks to maximize the difference between the scattering cross sections that the helix presents to the two circular polarizations of a single plane wave. The plane wave has a wavelength of 200 $\mu$m and its momentum is perpendicular to the helix axis. In this section, all the quantities are computed from the interaction matrices and are implicitly frequency (wavelength) dependent.

\begin{figure}[ht]
\includegraphics{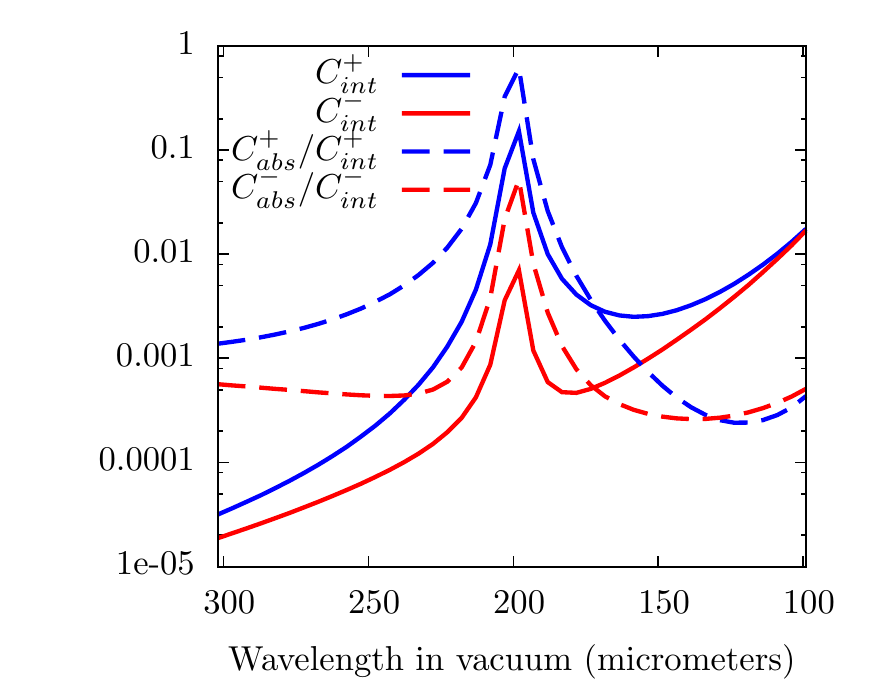}
\caption{\label{fig:lossfixed} Interaction cross sections and the ratio of absorption to interaction cross sections for each helicity $\pm$. At 200 $\mu$m, the maximum of em-chirality in Fig. (\ref{fig:chifixed}), there is non-negligible absorption of the helicity with dominant interaction cross section.}
\end{figure}

We now define two parameters which we use in the analysis. Recalling the definition of $C_{int}$ in Eq. (\ref{eq:cint}), we define the contrast $\Delta$ between the interaction cross sections corresponding to two input helicities as 
\begin{equation}
	\Delta=\frac{v_+^Tv_+-v_-^Tv_-}{C_{int}},
\end{equation}
and the total helicity change (duality breaking) $\cancel{D}$ as \footnote{This definition of $\cancel{D}$ is equivalent to the one in \cite[Eq.~(2)]{FerCor2015}.}
\begin{equation}
	\cancel{D}=\frac{\sigma\left(\Mpm\right)^T\sigma\left(\Mpm\right)+\sigma\left(\Mmp\right)^T\sigma\left(\Mmp\right)}{C_{int}}.
\end{equation}
The contrast $\Delta$ ranges from $-$1 to 1 and is equal to zero when the object presents the same interaction cross section to both helicities. The measure of duality breaking $\cancel{D}$ ranges from 0 to 1. Zero means complete helicity preservation and 1 complete helicity flipping.

Figure (\ref{fig:chifixed}) shows the normalized em-chirality $\chi/\sqrt{C_{int}}$, $\cancel{D}$, and $\Delta$ of the two turn helix as a function of the wavelength in vacuum. The maximum value of $\chi/\sqrt{C_{int}}=0.92$ is achieved near 200 $\mu$m. The figure illustrates the two conditions needed for large em-chirality: Large contrast between the two helicity interaction cross sections and small helicity change.

Figure (\ref{fig:lossfixed}) shows the interaction cross sections and the ratio of the absorption to the interaction cross sections for each input helicity. Interaction and absorption cross sections for each input helicity are defined in Eq. (\ref{eq:cintlambda}) and Eq. (\ref{eq:losses}), respectively. The figure shows that large values of em-chirality can be achieved in the presence of absorption losses. This is consistent with the fact that the conditions in Eqs.~(\ref{eq:mcrcons}) and (\ref{eq:mcrdipolar}) can be met by both lossy and lossless objects. This possibility originates in the use of reciprocity instead of time reversal invariance in Eq.~(\ref{eq:rec}), which avoids having to restrict the results to the lossless case. The principle of reciprocity has been recently used to obtain a theory of circular dichroism in planar systems \cite{Hopkins2015}.

The same design procedure that we followed at 200 $\mu$m produces significantly lower em-chirality values at higher frequencies. For example, at 20 $\mu$m the maximum of $\chi/\sqrt{C_{int}}$ that we obtain is equal to 0.65. While the design procedure that we used is not optimal, we take this as an indication that a different strategy and/or materials may be needed for maximizing em-chirality beyond the near infrared. As far as we know, objects with the desired properties are not yet available at optical or near UV frequencies, where they would be relevant for the two applications sketched in Sec. \ref{sec:applications}. We hope that our contribution increases the research in that direction. At optical frequencies, one may consider the fashioning of structures out of high index dielectric spheres meeting the dipolar duality condition. This strategy has recently been used to design an object which exhibits optical activity in general scattering directions \cite{FerCor2015}. As can be seen in \cite[Fig. 2a]{FerCor2015}, the restriction to dipolar duality constraints the choice of the spheres permittivity and radius to a narrow region in such parameter space.

\section{Conclusion}
In summary, we have defined the electromagnetic chirality of an object based on how it interacts with fields of different helicities (polarization handedness). The definition leads to relativistically invariant scalar measures of electromagnetic chirality.  Physical considerations allow to choose a particular measure. We have shown that the electromagnetic chirality of an object has an upper bound. The upper bound is equal to the square root of the interaction cross section of the object. Any object that is transparent to all fields of one helicity attains the upper bound: It is maximally electromagnetically chiral. For reciprocal objects, the implication goes the other way as well: Any maximally electromagnetically chiral and reciprocal object must be transparent to all fields of one helicity. Additionally, any maximally electromagnetically chiral and reciprocal object must have electromagnetic duality symmetry, i.e. it does not change the helicity of the fields interacting with it. We have derived the restrictions that these extremal objects must meet in two settings: The dipolar approximation and the macroscopic Maxwell's equations. The restrictions in their polarizability tensors or material constitutive relations are precise requirements for the design of maximally electromagnetically chiral objects. Electromagnetic duality symmetry is one of them. We have sketched two applications that show that these theoretical results also have practical value. Numerical analysis shows that, at least in a narrow frequency band, a realistic structure can come very close (92\%) to being maximally electromagnetically chiral even in the presence of losses. The analysis is also an example of how the theoretically obtained requirements can be used to guide a practical design.  
\nocite{Jackson1998,Lancaster1985,Cameron2012,Bliokh2013}
\begin{acknowledgments}
	IFC wishes to warmly thank Dr. Mauro Cirio for reading the manuscript and providing valuable feedback, and Ms. Magda Felo for her help with the figures. We also thank Dr. Tilo Arens for his comments on the manuscript. M.F. acknowledges support by the Karlsruhe School of Optics \& Photonics (KSOP). We also gratefully acknowledge financial support by the Deutsche Forschungsgemeinschaft (DFG) through RO 3640/3-1. Finally, we acknowledge support by Deutsche Forschungsgemeinschaft and Open Access Publishing Fund of Karlsruhe Institute of Technology.
\end{acknowledgments}

\end{document}